\documentclass{elsart4}

\usepackage{graphicx}

\usepackage{amssymb,amsbsy,amsmath}

\newcommand {\telcomp}{Sr$_{14}$Cu$_{24}$O$_{41}$}
\newcommand{\op}[1]{%
    \fontdimen12\textfont3=2pt\fontdimen12\scriptfont3=1.4pt%
    \!\null\mathop{\vphantom{#1}\smash{#1}}\limits_{\sim}\null\!}
\newcommand{\xref}[1]{\protect\ref{#1}}
\newcommand{\figref}[1]{Fig.~\protect\ref{#1}}
\newcommand{\fmref}[1]{(\protect\ref{#1})}

\def\geap{\raisebox{-.6ex}{$\stackrel {>}{\sim}$}} 

\begin{document}
\begin{frontmatter}

\title{Exact diagonalization studies of doped Heisenberg spin rings}


\author{J\"urgen Schnack\corauthref{cor1} and Fatiha Ouchni}
\address{Universit\"at Osnabr\"uck, Fachbereich Physik,
D-49069 Osnabr\"uck, Germany}
\corauth[cor1]{Tel: ++49 541 969-2695; fax: -12695; Email: jschnack@uos.de}

\begin{abstract}
  Motivated by magnetization studies of the ``telephon number
  compound'' \telcomp\ we investigate doped Heisenberg spin
  rings by means of complete numerical diagonalization of a
  Heisenberg Hamiltonian that depends parametrically on hole
  positions. A comparison with experimental magnetization data
  reveals rather accurate information about the screened
  electrostatic interaction between the charged holes on the
  ring which appears to be astonishingly strong.
\end{abstract}

\end{frontmatter}

\section{Introduction}
\label{sec-1}

Substances hosting spin and charge degrees of freedom promise a
large variety of phenomena like magnetic and charge ordering,
metallic conductivity and superconductivity
\cite{Dag:RMP94,KCG:PRB01}.  The ``telephone number compound''
\telcomp\ contains two magnetic one-dimensional structures,
chains and ladders. At low temperatures the ladder subsystems is
magnetically inactive due to a large spin gap \cite{TME:PRB98}.
The stochiometric formula \telcomp\ suggests 6 holes per fomula
unit. We will assume that for the undoped compound all holes are
located in the chain subsystem, although this is experimentally
under discussion since recent XAS measurements suggest that at
room temperature some holes are located in the ladder subsystem
\cite{NMK:PRB00} whereas it is necessary to assume that all
holes are in the chain subsystem in order to explain neutron
scatterin data \cite{RBM:PRB99}. The coupling between the Copper
spins depends on their exchange pathway across no, one, or two
holes.

The magnetism of the chain subsystem (for various dopings) has
been investigated for instance by means of classical spin
dynamics \cite{HoS:PRE03,SPV:EPJB02} or by comparison with
spin-dimer models \cite{CBC:PRL96,ABC:PRB00,Klingeler:2003}.
Usually the influence of the electrostatic hole-hole repulsion
is neglected and a certain ground state configuration based on
qualitative arguments is assumed. For the undoped compound the
so-called dimer configuration is supposed to be the ground state
configuration since this symmetric configuration should minimize
the Coulomb repulsion.  In addition, low-temperature
susceptibility \cite{Klingeler:2003}, neutron scattering data
\cite{RBM:PRB99,MYK:PRB99} as well as thermal expansion
measurements \cite{ABC:PRB00} strongly support the existence of
a low-lying dimer configuration.

The purpose of this article is as follows.  We want to introduce
a Heisenberg model which parametrically depends on hole
positions. If the electrostatic hole-hole repulsion is included
such a model allows to evaluate all energy eigenvalues and
eigenstates (for small system size) and thus enables us to
evaluate thermodynamic properties as function of temperature,
magnetic field, and doping. Assuming certain exchange constants
we can investigate the influence of the electrostatic hole-hole
repulsion on ground state properties as well as on thermal
averages like the magnetization which include contributions of
low-lying spin-hole configurations.

\section{Model system}
\label{sec-2}

There are various ways to model a system of holes and spins,
among them are the (multiband) Hubbard model and classical spin
dynamics models. In this article the chain of holes and spins is
modeled in the following way. Each configuration $\vec{c}$ of
holes and spins defines a Hilbert space which is orthogonal to
all Hilbert spaces arising from different configurations. The
Hamilton operator of a certain configuration is of Heisenberg
type and depends parametrically on the actual configuration
$\vec{c}$, i.~e.
\begin{eqnarray}
\label{E-1}
\op{H}(\vec{c})
&=&
-
\sum_{u,v}\;
J_{uv}(\vec{c})\;
\op{\vec{s}}(u) \cdot \op{\vec{s}}(v)
\ .
\end{eqnarray}
This ansatz is similar to a simple Born-Oppenheimer description
where the electronic Hamiltonian (here spin Hamiltonian) depends
parametrically on the positions of the classical nuclei (here
hole positions). In \eqref{E-1} $J_{uv}(\vec{c})$ are the
respective exchange parameters. In this article three exchange
parameters are considered: $J=-67$~K is the antiferromagnetic
coupling across a hole ($J=-64$~K and $J=-70$~K have been also
considered), $J_\parallel=5.8$~K is the ferromagnetic coupling
across two holes. The values of these exchange parameters are in
accord with other theoretical and experimental investigations of
this compound \cite{Klingeler:2003}. For the ferromagnetic
coupling between spins not separated by a hole we are using
$J_{NN}=8.7$~K, which is similar to the coupling used in
Ref.~\cite{CBC:PRL96}. A recent mean field analysis suggests
that this exchange constant may be stronger \cite{Klingeler:PC}.

The electrostatic interaction between holes is modeled
by a potential energy
\begin{eqnarray}
\label{E-2}
V(\vec{c})
&=&
\frac{e^2}{4\pi\epsilon_0\,\epsilon_r\,r_0}
\frac{1}{2}\;
\sum_{u\ne v}\;
\frac{1}{|u-v|}
\ ,
\end{eqnarray}
where $r_0=2.75$~\AA\ is the distance between nearest neighbor
sites on the ring and $\epsilon_r$ is the dielectric constant.
Several attempts have been undertaken to estimate the dielectric
constant which yielded values for $\epsilon_r$ up to 30
\cite{CPP:PRL89,BaS:PRB94,EKZ:PRB97}. In related projects where
the exchange interaction of chain systems in cuprates is derived
from hopping matrix elements between different orbitals using a
Madelung potential the dielectric constant is found to be 3.3
\cite{MTM:PRB98A,MTM:PRB98B}.

For small systems all hole configurations can be considered and
the related spin Hamiltonians \fmref{E-1} can be diagonalized
completely. For 8 spins and 12 holes this amounts to 6310
distinct hole configurations and tiny Hilbert spaces of
dimension 256. For 12 spins and 18 holes the total number of
hole configurations is already too big to be considered
completely. Therefore, only the dimer configuration and
low-lying excitations with their respective degeneracies are
taken into account. It will turn out that the high degeneracy of
excited hole configurations plays an important role, since they
substantially contribute to observables at low temperature
although lying rather high in energy.

\section{Numerical simulation and results}
\label{sec-3}

As an application we investigate spin rings where 60~\% of the
chain sites are occupied by holes. The ground state hole
configuration for this compound is the so-called dimer
configuration. In order to obtain this result it is necessary to
consider the full Coulomb interaction.  If one for instance
tries to model the Coulomb interaction by a nearest neighbor
repulsion, then the ground state is given by an
antiferromagnetic chain and a cluster of the remaining holes,
irrespective how big the nearest neighbor interaction is. Even
the inclusion of next-nearest neighbors does not improve the
situation, the Coulomb interaction is still proportional to the
number of sites and may be overcome by the antiferromagnetic
binding.  We also find that the Coulomb interaction has to be
sufficintly strong in order to yield the dimer configuration as
the ground state.  The following examples shows that this is
indeed the case, the Coulomb interaction is only weakly
screened.

\begin{figure}[ht!]
\centering
\includegraphics[clip,width=60mm]{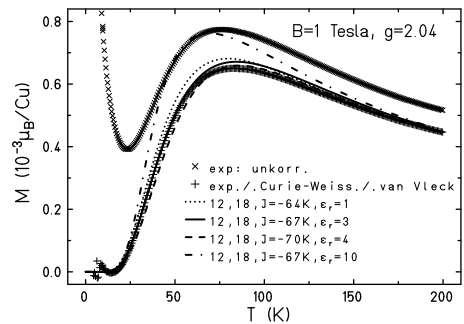}
\caption{Magnetization versus temperature for $B=1$~Tesla:
  Experimental data ($\vec{B}\parallel$ c-axis) are given by
  x-symbols and crosses (corrected for Curie-Weiss and van-Vleck
  magnetism) \cite{Klingeler:2003}. For $N_{\text{s}}=12$ and
  $N_{\text{h}}=18$ magnetization curves are displayed for a
  range of exchange parameters $J=-64, -67, -70$~K and several
  dielectric constants $\epsilon_r$.}
\label{F-C}
\end{figure}

Figure \xref{F-C} shows the temperature dependence of the
magnetization at $B=1$~Tesla for a system of 30 sites with
$N_{\text{s}}=12$ spins and $N_{\text{h}}=18$ holes. The
experimental data points for the infinite system
($\vec{B}\parallel$ c-axis) are given by x-symbols and crosses
(corrected for Curie-Weiss and van-Vleck magnetism)
\cite{Klingeler:2003}. The latter can be very accurately
approximated by an ansatz which describes the magnetic part as
noninteracting antiferromagnetically coupled dimers
\cite{Klingeler:2003}. Nevertheless, the additional
ferromagnetic coupling $J_\parallel=5.8$~K as well as other spin
hole configurations also contribute to the magnetization and
change the picture accordingly. If the Hamiltonian of the
12-spin system is diagonalized completely and Coulomb
interaction is taken into account, one finds that the
magnetization depends rather strongly on $J$ and $\epsilon_r$.
Figure \xref{F-C} displays magnetization curves for a range of
exchange parameters $J=-64, -67, -70$~K and several dielectric
constants $\epsilon_r$. For $\epsilon_r=1$ only the dimer
configuration contributes. For $\epsilon_r\geap 3$ several
hole configurations contribute with their respective magnetic
spectra. For practical purposes only those configurations have
been taken into account whose Coulomb energy differs by less
than 500~K from the ground state Coulomb energy.  It seems that
a dielectric constant of the order of $\epsilon_r\approx 3$ is
best suited to describe the magnetization data. This result is
in good agreement with a dielectric constant of 3.3 found in
Refs.  \cite{MTM:PRB98A,MTM:PRB98B}.

Figure \xref{F-D} illustrates which configurations contribute
for $\epsilon_r=3$ at lower energies. These configurations,
although rather high in energy, nevertheless contribute with
substantial weight, since they are highly degenerate. The
degeneracy for configurations where one hole is moved is of
order $N_{\text{tot}}$, the degeneracy for configurations where
two holes are moved is of order $N_{\text{s}}\cdot
N_{\text{tot}}$.

In order to estimate the influence of the finite size of the
ring system we computed the magnetization for a larger ring with
$N_{\text{s}}=16$ and $N_{\text{h}}=24$ which does not differ
from the respective one with $N_{\text{s}}=12$ and
$N_{\text{h}}=18$, therefore it seems to be justified to assume
that finite size effects do not play a role for the
magnetization at this size.

\section{Outlook}
\label{sec-4}

In this article a Heisenberg model which depends parametrically
on hole positions is introduced. It includes the electrostatic
repulsion between holes and allows to evaluate thermodynamic
properties as function of temperature, field, and doping. As a
first application the influence of the electrostatic repulsion
between holes on the temperature dependence of the magnetization
is investigated. We find that the dielectric constant is
approximately $\epsilon_r\approx 3$ in order to reproduce the
magnetization for 60~\% holes on the ring.

\begin{figure}[ht!]
\centering
\includegraphics[clip,width=60mm]{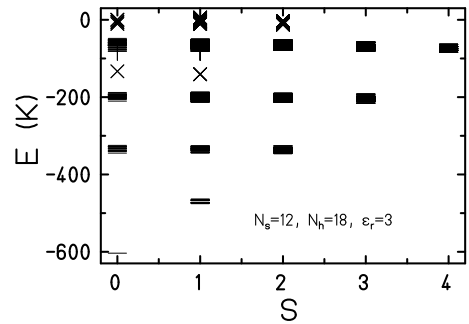}
\caption{Low-lying energy levels for $J=-67$~K and
  $\epsilon_r=3$: The dashes denote 
  magnetic levels of the dimer-configuration, x-symbols show
  levels for configurations where one hole is moved, the crosses
  mark levels for two holes moved. The first triplet excitation
  is split into four levels with the following excitation
  energies (and degeneracies): 128~K (3), 131~K (6), 136~K
  (6), and 139~K (3).}
\label{F-D}
\end{figure}

Since the proposed model depends at least on four parameters
($J, J_\parallel, J_{NN}, \epsilon_r$) the comparison to one
magnetization curve, \figref{F-C}, leaves some freedom for the
precise values. In future projects we are attempting to refine
the model in order to obtain a consistent description of the
meanwhile accumulated magnetization data. A direct measurement
of the energy needed to excite hole movements, compare
\figref{F-D}, would be very valuable since it would put
additional restrictions on the range of the dielectric constant
$\epsilon_r$. It is our hope that a refined model will allow
more insight into the interplay of charge order and magnetism.

\section*{Acknowledgement}

We would like to thank Bernd B\"uchner and especially R\"udiger
Klingeler (IFW Dresden) for valuable discussions and for the
supply with experimental data.



\end{document}